\documentclass[runningheads]{llncs}
\usepackage[most]{tcolorbox}
\usepackage{listings}
\usepackage{graphicx}
\usepackage{amssymb}
\usepackage{multirow}
\usepackage{todonotes}
\usepackage{esvect}
\usepackage{breqn}
\usepackage{breakcites}
\usepackage[tableposition=above]{caption}

\newcommand{\nextexperiment}{\refstepcounter{question}\arabic{question}}

\usepackage{hyperref,color}

\newcommand{\mstates}{S}
\newcommand{\mstate}{s}
\newcommand{\rstate}{r}

\newcommand{\symstate}{\sigma}
\newcommand{\initialsymbol}[1]{\mathbf{#1}}
\newcommand{\symexp}[1]{\mathbf{#1}}
\newcommand{\sympc}[1]{i_{#1}}
\newcommand{\sympath}[1]{\symexp{p}_{#1}}
\newcommand{\symmap}[1]{\symexp{m}_{#1}}
\newcommand{\symlist}[1]{\symexp{l}_{#1}}
\newcommand{\mapping}[2]{{#1} \rightarrow {#2}}

\newcommand{\birhalt}{$\texttt{HALT}$}

\newcommand{\bircjmp}[3]{$\texttt{CJMP}\ #1\ #2\ #3$}
\newcommand{\birobs}[2]{$\texttt{OBS}\ #1\ #2$}

\newcommand{\cons}{\circ}
\newcommand{\extag}[1]{\texttt{tag}(#1)}
\newcommand{\exidx}[1]{\texttt{index}(#1)}
\newcommand{\exoff}[1]{\texttt{off}(#1)}
\newcommand{\exobscond}[1]{\texttt{sline}(#1)}
\newcommand{\tool}{Scam-V}

\newcommand{\state}{s}

\newcommand{\smartparagraph}[1]{\smallskip \noindent \textbf{#1.}}

\makeatletter
\def\tcb@parbox@use@false{%
  \def\@parboxrestore{\linewidth\hsize\let\@parboxrestore=\tcb@parboxrestore}%
}
\makeatother
\usepackage[lighttt]{lmodern}

\lstdefinelanguage
   [x64]{Assembler}     
   {morekeywords={cbnz,ldr, cmp, ldp,b, eq, mul, nop, M, Z,%
                  udiv, bics, ldrsb, ror, %
                  xzr, wzr,x0,X1,x1,X2,x2,X3,x3,w4,x4,x5,x6,x7,x8,x9,%
                  x10,x11,x12,x13,x14,x15,x16,x17,x18,x19,%
                  x20,x21,x22,x23,x24,x35,x26,x27,x28,x29,%
                  x30,x31},
    alsoletter={x\#}}[strings,comments,keywords] 

\lstdefinestyle{blstenv}{
    language=[x64]Assembler,
    basicstyle=\footnotesize\ttfamily,
    keywordstyle=\bfseries,
    breaklines=true
    }
\newtcblisting{blstlisting}[2][]{    
    listing only, 
    listing style=blstenv,
    title=#2,
    #1
    }
\begin{document}

\title{Validation of Abstract Side-Channel Models for Computer Architectures}
%
%
\author{Hamed Nemati\inst{1} \and
Pablo Buiras\inst{2} \and
Andreas Lindner\inst{2} \and
Roberto Guanciale\inst{2}\and
Swen Jacobs\inst{1}}
\authorrunning{H. Nemati et al.}
%

\institute{Helmholtz Center for Information Security (CISPA), Saarbr{\"u}cken, Germany
\email{\{hnnemati, jacobs\}@cispa.saarland}\\
\and
KTH Royal Institute of Technology, Stockholm, Sweden\\
\email{\{buiras, andili, robertog\}@kth.se}}

\maketitle              
\begin{abstract}
Observational models make tractable the analysis of
information flow properties by providing an abstraction of side channels.
We introduce a methodology and a tool, \tool{}, to 
validate observational models for modern computer architectures.
We combine symbolic execution, relational analysis, and
different program generation techniques to generate experiments and validate the
models. An experiment consists of a randomly generated program together with two inputs
that are observationally equivalent according to the model under the test. Validation is done by 
checking indistinguishability of the two inputs on real hardware by
executing the program and analyzing the side channel.
We have evaluated our framework by validating models that abstract the data-cache side channel of
a Raspberry Pi 3 board with a processor implementing the ARMv8-A architecture. Our results show that \tool{} can identify
bugs in the implementation of the models and generate test programs which invalidate the models due to hidden microarchitectural
behavior.

\end{abstract}
%
%
%
%
%
%
%
%
%
%
%
\setcounter{footnote}{0} 

\section{Introduction}

Information flow analysis that takes into account side channels is a topic of increasing relevance, as attacks that compromise confidentiality via different microarchitectural features and sophisticated side channels continue to emerge~\cite{Kocher2018spectre,Osvik:2006:CAC:2117739.2117741,Gruss:2016:RRS:2976956.2976977,GuancialeNBD16,Kocher1996,Kocher:1999:DPA:646764.703989,aciiccmez2007predicting}. 
While there are information flow analyses that try to counter these threats~\cite{Almeida2016,Koepf2015}, these approaches use models that abstract from many features of
modern processors, like caches and pipelining, and their effects on channels that can be
accessed by an attacker, like execution time and power consumption.
Instead, these models~\cite{Molnar2006} include explicit ``observations'' that become available to an attacker when the program is executed and that should overapproximate the information that can be observed on the real system. 

While abstract models are indispensable for automatic verification because of the complexity of modern microarchitectures, the amount of details hidden by these models makes it hard to
trust that no information flow is missed, i.e., their soundness.
Different implementations of the same architecture, as well as optimizations such as parallel and
speculative execution, can introduce side channels that may be
overlooked by the abstract models.
This has been demonstrated by the
recent Spectre attacks~\cite{Kocher2018spectre}:
disregarding these microarchitectural features can lead to consider programs
that leak information on modern CPUs as secure.
Thus, it is essential to validate whether an abstract model
adequately reflects all information flows introduced by the low-level
features of a specific processor.

In this work, we introduce an approach that addresses this problem: we show 
how to validate observational models by comparing their 
outputs against the behavior of the real hardware in systematically generated 
experiments. 
In the following, we give an overview of our approach and this paper.

\begin{figure}[t]
\centering
  \includegraphics[width=0.85\linewidth]{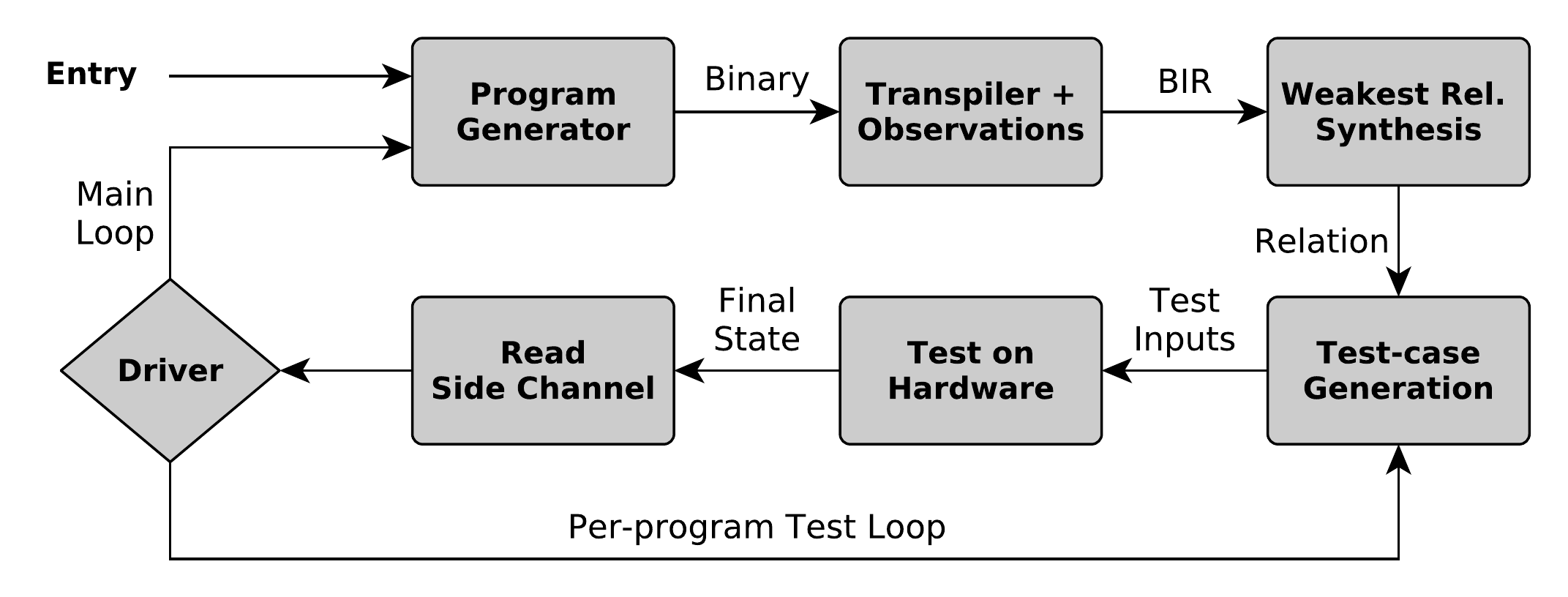}
  \caption{Validation framework workflow}
  \label{fig:arch}
\end{figure}

\smartparagraph{Our contribution}
We introduce \tool\ (Side Channel Abstract Model Validator), a framework for the automatic validation of abstract observational models.
At a high level,  \tool\ generates well-formed\footnote{ Terminating
  programs which do not cause run-time exceptions and emit
  observations required by the analysis.} random
 binaries and attempts to
construct pairs of initial states such that runs of the binaries from these states 
are indistinguishable at the
level of the model, but distinguishable on the
real hardware. In essence, finding such counterexamples implies that
the observational model is not sound, and leads to a potential
vulnerability. Fig.~\ref{fig:arch} illustrates the main workflow of \tool.

The first step of our workflow (described in 
Sect.~\ref{sec:program-generation}) 
is the generation of a binary program for the given architecture, 
guided towards programs that trigger certain
features of the architecture.
The second step translates the program to an intermediate
language (described in Sect.~\ref{sec:background:bir}) and 
annotates the result with observations according to
the observational model under validation. This transpilation is provably correct with
respect to the formal model of the ISA, i.e., the original binary program and the transpiled BIL program 
have the same effects on registers and memory.
In step three we use symbolic execution to synthesize the weakest relation on
program states that guarantees indistinguishability in the observational 
model (Sect.~\ref{sec:relation}). Through this relation, the observational model is used to drive the
generation  of \emph{test cases} -- pairs of states that satisfy the relation 
and can be used as inputs
to the program (Sect.~\ref{sec:test-generation}).
Finally, we run the generated binary with different test cases on the real 
hardware, and
compare the measurements on the side channel of the real processor.
A description of this process together with general remarks on our framework 
implementation
are in Sect.~\ref{sec:experiments}.
Since the generated test cases satisfy the synthesized relation,
soundness of the model would imply that the side-channel data on the real 
hardware cannot be distinguished
either. Thus, a test case where we can distinguish the two runs on the 
hardware amounts to a counterexample that invalidates the observational model.
After examining a given test case, the driver of the framework decides whether to 
generate more
test cases for the same program, or to generate a new program.

We have implemented \tool\ in the HOL4 theorem prover\footnote{\url{https://hol-theorem-prover.org}} and have evaluated the framework
on three observational models (introduced in Sect.~\ref{sec:background:obsmodels})
for the L1 data-cache of the ARMv8 processor on the
Raspberry Pi 3 (Sect.~\ref{sec:background:rpi}).
Our experiments (Sect.~\ref{sec:results}) led to the identification of model invalidating
microarchitectural features as well as bugs in the ARMv8 ISA
model and our observational extensions.
This shows that many existing abstractions are substantially unsound.

Since our goal is to validate that observational models
overapproximate hardware information flows, we do not attempt to
identify practically exploitable vulnerabilities. Instead, our
experiments attempt to validate these models in the worst case
scenario for the victim. This consists of an attacker that can
precisely identify the cache lines that have been evicted by the
victim and that can minimize the noise of these measurements
in the presence of background processes and interrupts.


\section{Background}\label{sec:background}
\subsection{Observational models}
We briefly introduce the concepts of side channels,
indistinguishability, observational models, and observational
equivalence.
For the rest of this section, consider a fixed program that runs on a fixed 
processor. We can model the program running on the processor by a transition 
system $M=\left<\mstates,\rightarrow\right>$, where $\mstates$ is a set of states 
and $\rightarrow \subseteq \mstates \times \mstates$ a transition relation.
In automated verification, the state space of such a model usually 
reflects the possible values of 
program variables (or: registers of the processor), abstracting 
from low-level behavior of the processor, 
such as cache contents, electric currents, or real-time behavior.
That is, for every state of the real system there is a state in the model 
that represents it, and a state of the model usually represents a set of 
states of the real system. 

Then, a \emph{side channel} is a trait of the real system that can be
read from by an attacker and that is not modeled in $M$.

\begin{definition}[Indistinguishability]\label{def:indistinguishable}
States $\rstate_1$  and $\rstate_2$ of the real system
are \emph{indistinguishable} if a real-world attacker is not able to
distinguish executions from $\rstate_1$ or $\rstate_2$  by means
of the side channel on the real hardware. 
\end{definition}
Note that executions may be
distinguishable even if they end in the same final state, e.g., if the attacker
is able to measure execution time.

In order to verify resilience against attacks that use side
channels, one option is to extend the model to include additional 
features of the real system
and to formalize indistinguishability in terms of some variations of
non-interference~\cite{GoguenM82a,GoguenM84}.
Unfortunately, it is infeasible to develop formal models that
capture \emph{all} side channels of a modern computer architecture.
For instance, precisely
determining execution time or power consumption of a program requires
to deal with complex processor features such as cache hierarchies, cache replacement policies, 
speculative execution, branch prediction, or bus arbitration.
Moreover, for some important parts of microarchitectures, their exact behavior may not even be public knowledge, e.g., the
mechanism used to train the branch predictor. Additionally,
information flow analyses
cannot use the same types of overapproximations that are used for
checking safety properties or analyzing worst-case execution time, e.g., the 
introduction of
nondeterminism to cover all possible outcomes.

In order to handle this complexity, information flow
analyses~\cite{Almeida2016, Koepf2015} use models designed to overapproximate 
information flow to channels in terms
of system state observations.
To this end, the model is extended with a set of possible observations $O$ 
and we consider a 
transition relation $\rightarrow \subseteq \mstates \times O \times \mstates$, 
i.e., 
each transition produces an observation that captures the 
information that it potentially leaks to the attacker.
We assume that the set $O$ contains an \emph{empty observation} $\perp$, and call a 
transition labeled with $\perp$ a \emph{silent} transition.
We call the resulting transition system an \emph{observational model}.
For instance, in case of a rudimentary cacheless processor, the
execution time of a program depends only on the sequence of
executed instructions. 
In this case, extending the model with observations that reveal the instructions
is more
convenient than producing a clock-accurate model of the system.

We use the operator $\circ$ for the sequential composition of observations. In 
particular, for 
a trace $\pi = \mstate_0 \rightarrow^{o_1} \mstate_1 \dots
\rightarrow^{o_n} \mstate_n$ of the model, we write 
$o_1 \circ \dots \circ o_n$ for the sequence of observations along $\pi$.
We write $o_1 \circ \dots \circ o_n \approx o_1' \circ \dots \circ o_{n'}'$ if the 
two sequences are equal after removing silent transitions.
Comparing traces with observations leads to a notion of
\emph{observational  equivalence}, defined as a relation on program
states.

\begin{definition}[Observational equivalence]\label{def:obsEq}
Traces $\pi = \mstate_0 \rightarrow^{o_1} \mstate_1 \dots
\rightarrow^{o_n} \mstate_n$ and 
$\pi' = \mstate'_0 \rightarrow^{o'_1} \mstate'_1 \dots
\rightarrow^{o'_{n'}} \mstate'_{n'}$
of an observational model $M$ are \emph{observationally equivalent} (written as $\pi \sim_M
\pi'$) iff $o_1 \circ \dots \circ o_n \approx o'_1 \circ \dots \circ
o'_{n'}$.

States $\mstate_1 \in \mstates$  and $\mstate_2 \in \mstates$ 
are \emph{observationally equivalent}, denoted $\mstate_1 \sim_M
\mstate_2$, iff for every possible trace $\pi_1$ of $M$ that starts in $\mstate_1$ there
is a trace $\pi_2$ of $M$ that starts in $\mstate_2$ such that $\pi_1 \sim_M \pi_2$, and vice versa.
\end{definition}

Note that this notion is, in principle, different from the
notion of \emph{indistinguishability}.
The overapproximation of information flows can lead to false
positives: for example, execution of a program may require the same
amount of time even if the sequences of executed instructions are
different.
A more severe concern is that these abstractions may overlook some flows of
information due to the number of low-level details that are hidden.
For instance, an observational model may not take into account that for some
microcontrollers the number of clock cycles required for
multiplication depends on the value of the operands.

The use of an abstract model to verify resilience against side-channel
attacks relies on the assumption that observational equivalence entails
indistinguishability for a real-world attacker on the real system:
\begin{definition}[Soundness]
  An observational model $M$ is \emph{sound} if whenever the model states $\mstate_1$ and $\mstate_2$ represent the real system states $\rstate_1$ and $\rstate_2$, respectively, then $\mstate_1 \sim_M
\mstate_2$ entails indistinguishability of $\rstate_1$ and $\rstate_2$.
\end{definition}

\subsection{The evaluation platform: Raspberry Pi 3}\label{sec:background:rpi}
In order to evaluate our framework, we selected Raspberry Pi 3\footnote{https://www.raspberrypi.org}, which is
a widely available ARMv8 embedded system.
The platform's CPU is a Cortex-A53, which is an 8-stage pipelined processor with
a 2-way superscalar and in-order execution pipeline. The CPU implements branch
prediction, but it does not support speculative execution. This makes
the CPU resilient against variations of Spectre attacks~\cite{a53spectre}.

In the following, we focus on side channels that exploit the Level 1 (L1) 
data-cache of the system.  The L1 data-cache is transparent for programmers. 
When the CPU needs to
read a location in memory in case of a cache miss, it copies the
data from memory into the cache for subsequent uses, tagging it with the 
memory location from which the data was read.
\begin{figure}[t]
\centering
\includegraphics[width=0.9\linewidth]{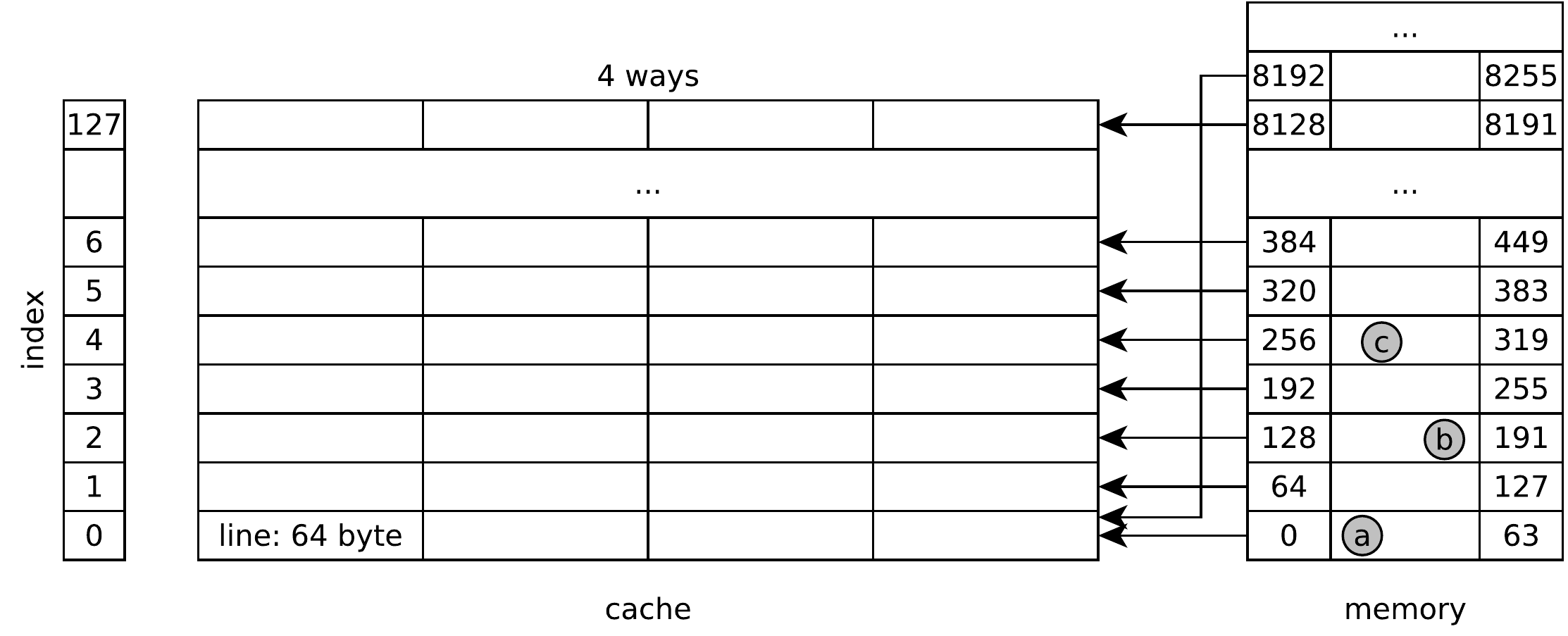}
\caption{L1 data-cache structure.}
\label{fig:datacache}
\end{figure}

Data is transferred between memory and cache in blocks of 64 bytes, called
cache lines.
The L1 data-cache (Fig.~\ref{fig:datacache}) is physically indexed and physically tagged and is 4-way set associative:
each memory location can be cached in four different entries in the
cache --- when a line is loaded, if all corresponding entries are
occupied, the CPU uses a specific (and usually underspecified)
replacement policy to decide which colliding line should be evicted.
The whole L1 cache is 32KB in size, hence it has 128 cache sets (i.e. 32KB / 64 B / 4). Let $a$ be a physical address, in the following we use
$\exoff{a}$ (i.e., least significant 6 bits),
$\exidx{a}$ (i.e., bits from 6 to 12), and
$\extag{a}$ (i.e., the remaining bits) to extract the
cache offset, cache set index, and cache tag of the address.

The cache implements a prefetcher, for some configurable $k \in \mathbb{N}$: when it detects a sequence of
$k$ cache misses
whose cache set indices are separated by a fixed stride, the
prefetcher starts to fetch data in the background.
For example, in Fig.~\ref{fig:datacache}, if $k=3$ and the cache is
initially empty
then accessing addresses $a$, $b$, and $c$, whose cache lines are
separated by a stride of 2,
can cause the cache to prefetch the block $[384 \dots 449]$.

\subsection{Different attacker and observational models}\label{sec:background:obsmodels}
Attacks that exploit the L1 data-cache are usually classified in three
categories:
In \emph{time-driven attacks} (e.g.~\cite{Tsunoo03cryptanalysisof}), the
attacker measures the execution time of the
victim and uses this knowledge to estimate the number of
cache misses and hits of the victim; In \emph{trace-driven attacks}
(e.g. ~\cite{Aciicmez:2006:TCA:2092880.2092891, zhang2012cross}), the adversary can profile the cache activities
during the execution of the victim
and observe the cache effects of a particular operation
performed by the victim; Finally, in \emph{access-driven attacks}
(e.g.~\cite{Neve:2006:AAC:1756516.1756531, Tromer:2010:ECA:1713125.1713127}), the
attacker can only determine the cache sets modified after the 
execution of the victim has completed.
A widely used approach to extract information via cache is
Prime+Probe~\cite{Osvik:2006:CAC:2117739.2117741}: 
(1) the attacker reads its own memory, filling the cache
with its data; (2) the victim is executed; (3) the attacker
measures the time needed to access the data loaded at step (1): slow
access means that the corresponding cache line has been evicted in
step (2). 

In the following we
disregard time-driven attacks and  trace-driven attacks: the
former can be countered by normalizing the victim execution time; the latter can be countered by preventing
victim preemption.
Focusing on
access-driven attacks leads to the following notion of
indistinguishability:
\begin{definition}\label{def:indistinguishablecache}
Real system states $\rstate_1$  and $\rstate_2$ 
are \emph{indistinguishable for access-driven attacks on the L1 data-cache}
 iff executions starting in $\rstate_1$ or
$\rstate_2$ modify the same cache sets.
\end{definition}

We remark that for multi-way caches, the need for models that overapproximate the
information flow is critical since the replacement policies are
seldom formally specified and a precise model of
the channel is not possible.
The following observational model attempts to overapproximate information flows
for data-caches by relying on the fact that accessing two different
addresses that only differ in their cache offset produces the same cache
effects:
\newcommand{\natnumbers}{\mathbb{N}}
\newcommand{\access}{\mathit{acc}}
\newcommand{\operation}{\mathit{op}}
\newcommand{\cidx}{i}
\newcommand{\ctag}{t}
\begin{definition}
  \label{model:multi-way-pc}
  The transition relation of the \emph{multi-way-pc cache observational
    model} is $\mstate \rightarrow_{mwc, pc}^o \mstate'$,
  where $\rightarrow_{mwc,pc}^o$ models the
 execution of one single instruction, with $o \in
  \natnumbers \times ((\{rd,wt\}
  \times \natnumbers \times \natnumbers)\ \cup \perp)$. 
  If $o=(pc, \access)$ then $pc$ is the current program counter
  and $\access=(\operation,\ctag,\cidx)$ is the memory access performed by the instruction,
  where $\operation$  is the memory operation, $\ctag$ is the cache tag and $\cidx$ is the cache set index
  corresponding to the address. If
  the instruction does not access the memory, then $\access=\perp$. 
\end{definition}
Notice that by making the program counter observable, this model
assumes that the attacker can infer the sequence of instructions
executed by the program. 

We introduce several relaxed models,
representing different assumptions on the hardware behavior and attacker
capability. Each relaxed model is obtained by projecting observations of
Def.~\ref{model:multi-way-pc}.
Let $\alpha$ be a relaxed model and $f_{\alpha}$ the corresponding projection function, 
then $\mstate \rightarrow_{\alpha}^{o'} \mstate'$ iff exists $o$ such
that $f_{\alpha}(o) = o'$ and $\mstate \rightarrow_{mwc, pc}^o \mstate'$.

The following model assumes that the
effects of instructions that do not interact with the data memory
are not measurable, hence the attacker does not observe the program counter:
\begin{definition}
  \label{model:multi-way}
  The projection of the \emph{multi-way cache observational
    model} is $f_{mwc}((pc, \access)) = \access$.
\end{definition}

On many processors, the replacement policy for a cache set does
not depend on previous accesses performed to other cache sets.
The resulting isolation among cache sets leads to the development of an
efficient countermeasure against access-driven attacks: 
cache coloring~\cite{godfrey2014preventing, 10.1145/325096.325161}. This consists in partitioning the cache sets into
multiple regions and ensuring that memory pages accessible by the
adversary are mapped to a specific region of the cache. In
this case, accesses to other regions do not affect the state of
cache sets that an attacker can examine. Therefore these accesses are not observable. This assumption is captured by the following model:
\begin{definition}
  \label{model:partitioning}
  The projection of the \emph{partitioned multi-way cache observational
    model} is $f_{pmwc}((pc, \access)) = \access$ if $\access = (\operation, \ctag, \cidx)$ and $\cidx$ belongs to
  the set of cache sets that are addressable by the attacker, and is
  $\perp$ otherwise.
\end{definition}
Notice that cache prefetching can violate soundness of this model,
since accesses to the non-observable region of the cache may lead to
prefetching addresses that lie in the observable part of the cache (see
Sect.~\ref{sec:results:part_cache}).

Finally, for direct-mapped caches,
where each memory address is mapped to only one cache entry, the cache tag
should not be observable if the attacker does not share memory with
the victim:
\begin{definition}
  \label{model:direct}
  The projection of the \emph{direct-mapped cache observational
    model} is $f_{dc}((pc, (\operation,\ctag,\cidx))) = (\operation,\cidx)$ and
  $f_{dc}((pc, \perp)) = \perp$.
\end{definition}
Since the cache in Cortex-A53 is multi-way set associative, this model is not sound. For example, 
in a two-way set associative cache, accessing $a, a$ and $a, b$, where
both $a$ and $b$ have the same cache set index but different cache tags, may result in
different cache states.

\subsection{Binary Intermediate Representation}\label{sec:background:bir}
\begin{figure}[t]
\begin{blstlisting}[hbox,enhanced,titlerule = 1pt,boxsep = 1pt,left = 10pt,right = 10pt,top=2pt,colframe = gray,center,parbox=false]{BIR program with observation}
//b.eq l2
[l0:CJMP Z l2 l1]
//mul x1 x2 x3
[l1:X1= X2*X3; JMP l2]
//ldr x2 {x1} +8
[l2:OBS(sline(X1),[tag(X1),index(X1)]);
    X2= LOAD(M, X1);X1= X1+8;HALT]
\end{blstlisting}
\caption{BIR transpilation example}
\label{fig:bir_transpiled}
\end{figure}

To achieve a degree of hardware independence, we
use an architecture-agnostic intermediate representation
known as BIR~\cite{Lindner2019}. BIR is an abstract assembly language 
with statements that work on memory, arithmetic expressions, and 
jumps. Fig.~\ref{fig:bir_transpiled}
shows an example of code in a generic assembly language and its transpiled BIR code.
This code performs a conditional jump to \textit{l2} if \textbf{Z} holds,
and otherwise it sets $\initialsymbol{X1}$ to the multiplication $\initialsymbol{X2 * X3}$. Then,
at \textit{l2} it loads a word from memory at address $\initialsymbol{X1}$ into
$\initialsymbol{X2}$, and finally adds 8 to the pointer $\initialsymbol{X1}$. BIR programs are organized into blocks,
which consist of jump-free statements and end in either conditional jump (\texttt{CJMP}),
unconditional jump (\texttt{JMP}), or \texttt{HALT}.

BIR also has explicit support for \emph{observations}, which are produced by
statements that evaluate a list of expressions in the current state.
To account for
expressive observational models, BIR allows conditional observation. The condition
is represented by an expression attached to the observation statement. The
observation itself happens only if this condition evaluates as true in the
current state. The observations in Fig.~\ref{fig:bir_transpiled} reflect a scenario where the 
data-cache has been partitioned: some lines are exclusively
accessible by the victim (i.e. the program), some lines can be shared
with the attacker.
The statement
\texttt{OBS(sline$(\initialsymbol{X1}$),[tag($\initialsymbol{X1}$),
  index($\initialsymbol{X1}$)])} for the load instruction consists of an
observation condition (\texttt{sline($\initialsymbol{X1}$)}) and a list of expressions to observe
(\texttt{[tag($\initialsymbol{X1}$), index($\initialsymbol{X1}$)]}).
The function \texttt{sline} checks that the argument address is mapped
in a shared line and therefore visible to the attacker. The
functions \texttt{tag} and \texttt{index}
extract the cache tag and set index in which the argument address is mapped.
Binary programs can be translated to BIR
via a process called \emph{transpilation}. This
transformation reuses formal models of the ISAs
and generates a proof that certifies
correctness of the translation by establishing a bisimulation between
the two programs.


\newcommand{\type}[1]{\ensuremath{\mathit{#1}}}
\newcommand{\term}[1]{\ensuremath{\mathtt{#1}}}
\newcommand{\bind}[0]{\ensuremath{\mathtt{>\!\!>\!=}}}

\section{Program generation}
\label{sec:program-generation}

We base our validation of observational models on the execution of binary programs rather than higher-level code representations.
This approach has the following benefits:
(i) It obviates the necessity to trust compilers or reason about how their compilation affects side-channels.
(ii) Implementation effort is reduced because most existing side-channel analysis approaches also operate on binary representations, which requires ISA models.
(iii) This approach allows to find ISA model faults independently of the compilation.
(iv) It enables a unified infrastructure to handle many different types of channels.

In \tool{}, we implemented
two techniques to generate well-formed binaries: \emph{random} program generation and \emph{monadic}
program generation. The random generator leverages the instruction encoding
machinery from the existing HOL4 model of the ISA and produces arbitrary well-formed ARMv8
binaries, with the possibility to control the frequency of occurrences of each
instruction class. The monadic generator is following a grammar-driven approach in the
style of QuickCheck~\cite{Claessen00} that generates arbitrary programs that fit
a specific pattern or template. The program templates can be defined in a
modular, declarative style and are extensible. We use this approach to generate
programs in a guided fashion, focusing on processor features that we want to
exercise in order to validate a model, or those we suspect may lead to a
counterexample. Fig.s~\ref{fig:proggen_example_randprog_cbnz}~and~\ref{fig:proggen_examples} show some example programs generated by \tool,
including straight-line programs that only do memory loads, programs that load
from addresses in a stride pattern to trigger automatic prefetching, and
programs with branches. 
More details on how the program generators work can be found in Appendix~\ref{app:proggen}.
{
  \begin{figure}[t]
    \centering
  \begin{blstlisting}[hbox,scale=0.9,enhanced,titlerule = 1pt,boxsep = 1pt,left = 10pt,right = 10pt,top=2pt,colframe = gray,center,parbox=false]{Random program generator}
udiv x3, x16, x8
cbnz x28, #12
bics x14, x3, x26, ror #21
ldrsb w4, [x24, x3]
ldp x22, x14, [x3], #0xD0
\end{blstlisting}
\caption{Example programs generated by the \tool{} random program generator.}
\label{fig:proggen_example_randprog_cbnz}
\end{figure}
}

{
  \setlength{\belowcaptionskip}{-10pt}
  \begin{figure}[t]
    \centering
  \begin{blstlisting}[hbox,scale=0.9,enhanced,titlerule = 1pt,boxsep = 1pt,left = 10pt,right = 10pt,top=2pt,colframe = gray,center,parbox=false]{Load generator\hspace{2cm}Load strides\hspace{2cm}Load seq. with branches}
ldr x5, [x1, 4] | ldr x16, [x2, #0]    | cmp x13, x3            
ldr x0, x3      | ldr x19, [x2, #64]   | b.eq #0x14
ldr x2, [x9, 8] | ldr x1,  [x2, #128]  | ldr x6, #0x1DFA
                | ldr x22, [x2, #192]  | ldr x9, [x20]
                | ldr x17, [x2, #256]  | ldr x20, [x22, #8]
                |                      | b #0x10
                |                      | ldr x16, [x0, #16]
                |                      | ldr x16, #0x1BE8
                |                      | ldr x12, #0x17D5
\end{blstlisting}
\caption{Example programs generated by \tool{} monadic program generators.}
\label{fig:proggen_examples}
\end{figure}
}


\section{Synthesis of Weakest relation}
\label{sec:relation}
Synthesis of the weakest relation is based on standard symbolic
execution techniques.
We only cover the basic ideas of symbolic execution in the following
and refer the reader to \cite{10.1145/360248.360252} for more details.
We use $\initialsymbol{X}$ to range over symbols,
and $\symexp{c}$, $\symexp{e}$, and
$\symexp{p}$ to range over symbolic expressions.
A symbolic state $\symstate$ consists of a concrete program counter
$\sympc{\symstate}$,
a path condition $\sympath{\symstate}$,
and a mapping $\symmap{\symstate}$ from variables to symbolic
expressions.
We write $e(\symstate) = \symexp{e}$ for the
symbolic evaluation of the expression $e$ in $\symstate$,  
and $\symexp{e}(s)$ for 
the value obtained by substituting the symbols of the symbolic
expression $\symexp{e}$ with the values of the variables in $s$, where $s$ is a concrete state.

Symbolic execution produces one terminating
state\footnote{We consider only terminating programs.}
for
each possible execution path: a terminating
state is produced when \birhalt\ is encountered; the execution of
\bircjmp{c}{l_1}{l_2} from state $\symstate$ follows both branches using the
path conditions $c(\symstate)$ and $\neg c(\symstate)$. 
Symbolic execution of the example in Fig.~\ref{fig:bir_transpiled}
produces the terminating states $\symstate_1$ and $\symstate_2$. For the first branch we have  
 $\sympath{\symstate_1} = \initialsymbol{Z}$ and
$\symmap{\symstate_1} = 
\{
\mapping{X_1}{\initialsymbol{X}_1+8},
\mapping{X_2}{\texttt{LOAD}(\initialsymbol{M}, \initialsymbol{X}_1)}
\}
$ (we omit the variables that are not updated),
and for the second branch 
$\sympath{\symstate_2} = \neg \initialsymbol{Z}$ and
$\symmap{\symstate_2} = 
\{
\mapping{X_1}{\initialsymbol{X}_2 * \initialsymbol{X}_3 + 8},
\mapping{X_2}{\texttt{LOAD}(\initialsymbol{M}, \initialsymbol{X}_2 * \initialsymbol{X}_3)}
\}
$.

\newcommand{\explist}[1]{\vv{#1}}

We extend standard symbolic execution to handle observations. That is,
we add to each symbolic state a list $\symlist{\symstate}$, and the execution of \birobs{c}{\explist e} in $\symstate$
appends the pair $(\symexp{c}, \explist
{\symexp{e}})$ to $\symlist{\symstate}$,
where $\symexp{c} = c(\symstate)$ and $\explist {\symexp{e}}[i] = \explist
{e}[i](\symstate)$ are the symbolic evaluation of the condition and
expressions of the observation.
For instance, in   
the example of Fig.~\ref{fig:bir_transpiled} the list for the terminating states are
\[
  \begin{array}{ll}
    \symlist{\symstate_1} & = [(\exobscond{\initialsymbol{X}_1},
                            [\extag{\initialsymbol{X}_1}, \exidx{\initialsymbol{X}_1}])]\\
    \symlist{\symstate_2} & = [
                            (\exobscond{\initialsymbol{X}_2 *\initialsymbol{X}_3},
                            [\extag{\initialsymbol{X}_2 *
                            \initialsymbol{X}_3},
                            \exidx{\initialsymbol{X}_2 * \initialsymbol{X}_3}])]
  \end{array}
\]
Let $\Sigma$ be the set of terminating states produced by the symbolic
execution, $s$ be a concrete state, and 
$\symstate \in \Sigma$
be a symbolic state such that $\sympath{\symstate}(s)$ holds, then
executing the program from the initial state $s$ produces the 
value $\symmap{\symstate}(X)(s)$ for the variable $X$.
Moreover, let
$\symlist{\symstate} = [(\symexp{c}_1, {\explist {\symexp{e}}_1}) \dots
(\symexp{c}_n, {\explist {\symexp{e}}_n})]$,
then the generated observations are $(\symexp{c}_1, {\explist {\symexp{e}}_1})(s)
\cons \dots \cons (\symexp{c}_n, {\explist {\symexp{e}}_n})(s)$,
where 
$(\symexp{c}_1, {\explist {\symexp{e}}_1})(s) = {\explist
  {\symexp{e}}}_1(s)$ if $\symexp{c}_1(s)$, and otherwise $\perp$
(i.e. observations are list of concrete values).

After computing $\Sigma$, we synthesize the observational equivalence
relation (denoted by $\sim$)  by
ensuring that every possible pair of execution paths have equivalent 
lists of observations. 
Formally, $s_1 \sim s_2$ is equivalent to:
\[
\bigwedge_{(\symstate_1, \symstate_2) \in \Sigma \times \Sigma}
(\sympath{\symstate_1}(s_1) \land \sympath{\symstate_2}(s_2) \Rightarrow
\symlist{\symstate_1}(s_1) = \symlist{\symstate_2}(s_2))
\]
\newcommand{\exobsvaleq}[2]{(\extag{#1} = \extag{#2}  \land \exidx{#1} = \exidx{#2})}
\newcommand{\exobstwo}[2]{\exobscond{#1} \land \exobscond{#2} \Rightarrow (\exobsvaleq{#1}{#2})}
This synthesized relation implies the observational equivalence defined in Sect.~\ref{sec:background} (Def.~\ref{def:obsEq}). In the example, the synthesized relation (after simplification) is as
follows (notice that primed symbols represent variables of the second
state and we omitted the symmetric cases):
\begin{dmath*}
 \fontsize{7.8}{9}
  \begin{array}{ll}
   (\initialsymbol{Z}\; \land\; \initialsymbol{Z'})\; \Rightarrow \\
    \;\;\left(\begin{array}{l}
    \exobscond{\initialsymbol{X}_1} = \exobscond{\initialsymbol{X'}_1}\; \land\\ \exobscond{\initialsymbol{X}_1} \Rightarrow \exobsvaleq{\initialsymbol{X}_1}{\initialsymbol{X'}_1}\\
    \end{array}\right)\;\; \land\\

    (\initialsymbol{Z} \land \neg \initialsymbol{Z'}) \Rightarrow \\
    \;\;\left(\begin{array}{l}
        \exobscond{\initialsymbol{X}_1} = \exobscond{\initialsymbol{X'}_2 * \initialsymbol{X'}_3}\; \land \\    
        \exobscond{\initialsymbol{X}_1} \Rightarrow
        \exobsvaleq{\initialsymbol{X}_1}{\initialsymbol{X'}_2 * \initialsymbol{X'}_3}\\
    \end{array}\right)\;\; \land\\ 
    
    (\neg \initialsymbol{Z} \land\! \neg \initialsymbol{Z'}) \Rightarrow \\
    \;\;\left(\begin{array}{l}
        \exobscond{\initialsymbol{X}_2\! *\! \initialsymbol{X}_3} = \exobscond{\initialsymbol{X'}_2\! *\! \initialsymbol{X'}_3}\; \land \\
        \exobscond{\initialsymbol{X}_2\! *\! \initialsymbol{X}_3} \Rightarrow
        \exobsvaleq{\initialsymbol{X}_2\! *\! \initialsymbol{X}_3}{\initialsymbol{X'}_2\! *\! \initialsymbol{X'}_3}
    \end{array}\right)
  \end{array}
\end{dmath*}

\begin{figure}[t]
\centering
$
  \begin{array}{cccccc}
    s_1= & &  s_2 =  & s'_1= & &  s'_2 = \\
    \left \{
    \begin{array}{ll}
      \initialsymbol{Z} = T\\
      \initialsymbol{X}_1 = 130\\
      \initialsymbol{X}_2 = 123546\\
      \initialsymbol{X}_3 = 87465
    \end{array}
    \right \} & \sim &
    \left \{
    \begin{array}{ll}
      \initialsymbol{Z} = F\\
      \initialsymbol{X}_1 = 37846\\
      \initialsymbol{X}_2 = 2\\
      \initialsymbol{X}_3 = 64
    \end{array}
    \right \}
    \hspace{0.5em} &
    \left \{
    \begin{array}{ll}
      \initialsymbol{Z} = F\\
      \initialsymbol{X}_1 = 3246\\
      \initialsymbol{X}_2 = 64\\
      \initialsymbol{X}_3 = 30
    \end{array}
    \right \} & \sim &
    \left \{
    \begin{array}{ll}
      \initialsymbol{Z} = F\\
      \initialsymbol{X}_1 = 856\\
      \initialsymbol{X}_2 = 12\\
      \initialsymbol{X}_3 = 64
    \end{array}
    \right \}

  \end{array}
$
\caption{\label{fig:testcases} Example test cases when the first 10 cache sets are shared.}
\end{figure}

We recall that Raspberry Pi 3 has 128 cache sets and 64 bytes per
line.
Fig.~\ref{fig:testcases} shows two pairs of states that satisfy the relation,
assuming only the first 10 cache sets are shared.
States $s_1$ and $s_2$ lead the program to access the third cache set, while
$s'_1$ and $s'_2$ lead the program to access cache sets
that are not shared, therefore they generate no observations.


\section{Test-Case Generation}
\label{sec:test-generation}

A test case for a program $P$ is a pair of initial states $\state_1$, $\state_2$ such that
$P$ produces the same observations when executed from either state,
i.e., $\state_1 \sim \state_2$. The relation as described in
Sect.~\ref{sec:relation} characterizes the space of observationally equivalent
states, so a simple but naive approach to test-case generation consists in
querying the SMT solver for a model of this relation.
The model that results from the query gives us two concrete
observationally equivalent values for the registers
that affect the observations
of the program, so at this point we could forward these to our testing
infrastructure to perform the experiment on the hardware.

However, the size of an observational equivalence class can be enormous, because
there are many variations to the initial states that cannot have effects on the
channels available to the attacker. Choosing a satisfying assignment for the
entire relation every time without any extra guidance risks producing many test
cases that are too similar to each other, and thus unlikely to find
counterexamples. For instance, the SMT solver may generate many
variations of the test case $(s_1,s_2)$ in Fig.~\ref{fig:testcases} by iterating over all
possible values for register $X_2$ of state $\state_1$, even if the
value of this register is immaterial for the observation.

In practice, we explore the space of observationally equivalent
states in a more systematic manner.
To this end, \tool{} supports two mechanisms to guide the selection of test
cases: \emph{path enumeration} and \emph{term enumeration}. Path enumeration
partitions the space according to the combination of symbolic execution paths
that are taken, whereas term enumeration partitions the space according to the value
of a user-supplied BIR expression. In both cases, the partitions are explored in
round-robin fashion, choosing one test case from each partition in turn. To make
the queries to the SMT solver more efficient, we only generate a fragment of the
relation that corresponds to the partition under test.

\smartparagraph{Path enumeration} Every time we have to generate a test case, we
first select a pair $(\symstate_1, \symstate_2) \in \Sigma \times
\Sigma$ of symbolic states as per Sect.~\ref{sec:relation}, which
identifies a pair of paths 
$(\sympath{\symstate_1},
\sympath{\symstate_2})$. The
chosen paths vary in each iteration in order to achieve full path coverage.
The query given to the SMT solver then becomes\footnote{Note that this is equivalent to taking a fragment of the observational
equivalence relation, specifically the case when $  \sympath{\symstate_1}(\state_1)
\land \sympath{\symstate_2}(\state_2)$ holds.}
\[
  \sympath{\symstate_1}(\state_1) \land \sympath{\symstate_2}(\state_2) \land
  \symlist{\symstate_1}(\state_1) = \symlist{\symstate_2}(\state_2)
\]

Since the meat of the relation is a conjunction of implications, this is a natural
partitioning scheme that ensures all conjuncts are actually explored. Note that
without this mechanism, the SMT solver could always choose states that only
satisfy one and the same conjunct.
To guide this process even further, the user can supply a \emph{path guard},
which is a predicate on the space of paths. Any path not satisfying the
guard is skipped, allowing the user to avoid exploring unwanted paths.
For example, for the program in Fig.~\ref{fig:bir_transpiled} we can use a path guard to
force the test generation to select only paths that produce no
observations: e.g.,
 $ (\initialsymbol{Z} \Rightarrow
    \neg \exobscond{\initialsymbol{X}_1})  \land
   (\neg \initialsymbol{Z} \Rightarrow
    \neg \exobscond{\initialsymbol{X}_2 * \initialsymbol{X}_3})
$.

\smartparagraph{Term enumeration} In addition to path enumeration, we can choose a
BIR expression $e$ that depends on the symbolic state, and a range $R$ of values
to enumerate. Every query also includes the conjuncts $e_{\symstate_1} = v_1 \land
e_{\symstate_2} = v_2$ where $v_1,v_2\in R$ and such that the $v_i$ are chosen
to achieve full coverage of $R \times R$. Term enumeration can be useful to
introduce domain-specific partitions, provided that $R\times R$ is small enough.
For example, this mechanism can be used to ensure that we explore addresses that
cover all possible cache sets, if we set $e$ to be a mask that extracts the
cache set index bits of the address. For example, for the program in Fig.~\ref{fig:bir_transpiled}
we can use
  $     \initialsymbol{Z}  * \exidx{\initialsymbol{X}_1} +
   (1 - \initialsymbol{Z}) * \exidx{\initialsymbol{X}_2 * \initialsymbol{X}_3}
$ to enumerate all combinations of accessed cache sets while
respecting the paths.



\section{Implementation}\label{sec:experiments}
The implementation
\footnote{
Our implementation of \tool{} is embedded in HolBA, which is available at \url{https://github.com/kth-step/HolBA}.
Our extendable experimentation platform consists of several "EmbExp-*" repositories available at \url{https://github.com/kth-step}.
}
of \tool{} is done in the HOL4 theorem
prover using its meta-language, i.e., SML. \tool{} relies
on the binary analysis platform HolBA
for transpiling the binary code of test programs to the BIR
representation.
This transpilation uses the existing HOL4 model of the ARMv8
architecture~\cite{FoxL3modelsweb} for giving semantics to ARM programs.
In order to validate the observational models of
Sect.~\ref{sec:background:obsmodels}, we extended the transpilation 
process to inline observation statements into the resulting BIR
program. These observations represent the observational power of the side channel.
In order to compute possible execution paths of test programs and
their corresponding observations, which are needed to synthesize the observational
equivalence relation of Sect.~\ref{sec:relation}, we implemented a symbolic execution engine in HOL4.
All program generators from Sect.~\ref{sec:program-generation} as well as
the weakest relation synthesis from Sect.~\ref{sec:relation} and
the test-case generator from Sect.~\ref{sec:test-generation} are implemented
as SML libraries in \tool{}. The latter uses the SMT solver Z3~\cite{10.5555/1792734.1792766} to generate test inputs.
For conducting the experiments in this paper, we used
Raspberry Pi 3 boards equipped with ARM Cortex-A53
processors implementing the ARMv8-A architecture.

\begin{figure}[t]
\centering
\includegraphics[clip, trim=0.7cm 0.5cm 0.7cm 0.5cm, width=0.8\linewidth]{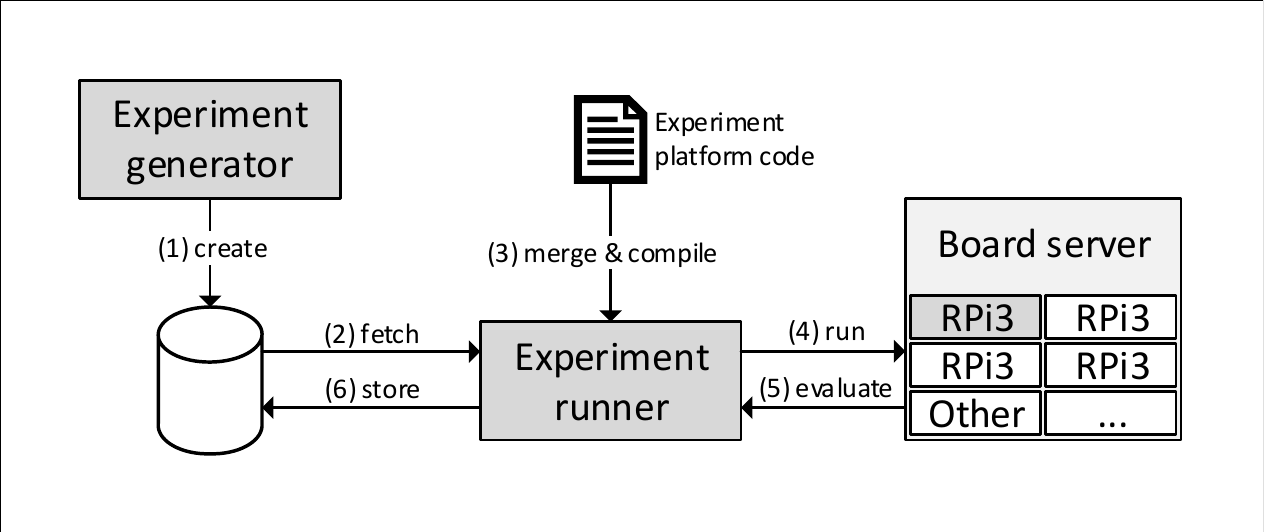}
\caption{Experiment handling design with numbered steps. This showcases the workflow for producing, preparing, executing and evaluating one experiment.}
\label{fig:experiment_arch}
\end{figure}

The \tool\ pipeline generates programs and pairs of observationally equivalent
initial states (test cases) for each program. Each combination of a program with
one of its test cases is called an \emph{experiment}.
After generating experiments, we execute them on the processor implementation of interest to examine their effects on the side channel. 
Fig.~\ref{fig:experiment_arch} depicts the life of a single experiment as goes
through our experiment handling design. This
consists of: (step 1) generating an experiment and storing it in a database, (step
2) retrieving the experiment from the database, (step 3) integrating it with \textit{experiment-platform code}
and compiling it into platform-compatible machine code, and (step
4-6) executing the generated binary on the real board, receiving and
storing the experiment result.

The experiment-platform code 
configures page tables to setup \textit{cacheable} and
\textit{uncacheable} memory, clears the cache before every execution
of the program, and inserts memory barriers around the experiment code.
The platform executes in ARM TrustZone, which enables us to use privileged debug
instructions to obtain the cache state directly for comparison after experiment execution.

The way in which we compare final cache states for distinguishability depends on
the attacker and observational model in question. For multi-way cache, 
we say two states are \emph{indistinguishable} if and only if for each valid entry in one state, there is a valid entry with the same cache tag in the corresponding cache set of the other state and vice versa.
For the partitioned multi-way cache, we check the
states in the same way, except we do it only for a subset of the cache sets (see
Sect.~\ref{sec:results:part_cache} for details on the exact partition). For
the direct-mapped cache, we compare how many valid cache lines there are in each
set, disregarding the cache tags. These comparison functions have been chosen to match
the attacker power of the relaxed models in Def.s~\ref{model:multi-way},
\ref{model:partitioning}, and \ref{model:direct} respectively.


\section{Results}\label{sec:results}
Since the ARM-v8 experimentation platform runs as bare-metal code,
there are no background processes or interrupts.
Despite this fact,
our measurements may contain noise due to other hardware components
that share the same memory subsystem, such as the GPU, and because our
experiments are not synchronized with  
the memory controller.
In order to
simplify repeatability of our experiments, we execute each experiment
10 times and check for discrepancies in the final state of the data
cache. Unless all executions give the same result, this experiment is
classified as \textit{inconclusive} and excluded from further analysis.

\subsection{Direct-mapped cache observational model}
First, we want to make sure that \tool{} can invalidate unsound observational
models in general. For this purpose, we generated experiments that use the model of
Def.~\ref{model:direct}, i.e., for every memory 
access in BIR we observe the cache set index of the address of the access.
We know that this is not a sound model for Raspberry Pi 3, because the platform uses a 4-way
cache.
Table~\ref{table:exps_res}.\ref{table:exps_obsindexonly} shows that both the random program
generator and the monadic load generator uncovered
counterexamples that invalidated this observational model.

\begin{table}[t]
\centering
\begin{tabular}{|c|l|l|l|l|}
\hline
\multirow{5}{*}{(\ref{table:exps_res}.\nextexperiment\label{table:exps_obsindexonly})} & Observations            & \multicolumn{3}{l|}{Cache set index only (Def.~\ref{model:direct})}                                                                                                                                                                        \\ \cline{2-5} 
                     & Programs                                                                                & Monadic load generator                                                                                   & \multicolumn{2}{l|}{Random program generator}                                                                  \\ \cline{2-5} 
                     & \begin{tabular}[c]{@{}l@{}}Experiments\\ - Inconclusive\\ - Counterexample\end{tabular} & \begin{tabular}[c]{@{}l@{}}39660\\ 0\\ 19\end{tabular}                                                   & \multicolumn{2}{l|}{\begin{tabular}[c]{@{}l@{}}20872\\ 1\\ 18\end{tabular}}                                    \\ \hline\hline
\multirow{7}{*}{(\ref{table:exps_res}.\nextexperiment\label{table:exps_cachepart})} & Experiment set              & \multicolumn{1}{l|}{Variation A}                                                                         & \multicolumn{2}{l|}{Variation B}                                                                               \\ \cline{2-5} 
                     & Observations                                                                            & \multicolumn{1}{l|}{\begin{tabular}[l]{@{}c@{}}Page unaligned cache \\ partitioning (Def.~\ref{model:partitioning})\end{tabular}} & \multicolumn{2}{l|}{\begin{tabular}[l]{@{}l@{}}Page aligned cache\\  partitioning (Def.~\ref{model:partitioning})\end{tabular}}         \\ \cline{2-5} 
                     & Programs                                                                                & \multicolumn{3}{l|}{Monadic stride generator}                                                                                                                                                                             \\ \cline{2-5} 
                     & \begin{tabular}[c]{@{}l@{}}Experiments\\ - Inconclusive\\ - Counterexample\end{tabular} & \begin{tabular}[c]{@{}l@{}}36160\\ 5426\\ 3460\end{tabular}                                              & \multicolumn{2}{l|}{\begin{tabular}[c]{@{}l@{}}37843\\ 6967\\ 0\end{tabular}}                                  \\ \hline\hline
\multirow{6}{*}{(\ref{table:exps_res}.\nextexperiment\label{table:exps_cachetagandsetindex})} & Observations      & \multicolumn{3}{l|}{Cache tag and set index (Def.~\ref{model:multi-way})}                                                                                                                                                                              \\ \cline{2-5} 
                     & \multirow{2}{*}{Programs}                                                               & \multirow{2}{*}{Random program generator}                                                                & \multicolumn{2}{c|}{Monadic generator}                                                                         \\ \cline{4-5} 
                     &                                                                                         &                                                                                                          & Loads \hspace{10mm}                                   & Previction                                             \\ \cline{2-5} 
                     & \begin{tabular}[c]{@{}l@{}}Experiments\\ - Inconclusive\\ - Counterexample\end{tabular} & \begin{tabular}[c]{@{}l@{}}20256\\ 2\\ 0\end{tabular}                                                    & \begin{tabular}[c]{@{}l@{}}23120\\ 0\\ 5\end{tabular} & \begin{tabular}[c]{@{}l@{}}23290\\ 0\\ 16\end{tabular} \\ \hline\hline
\multirow{5}{*}{(\ref{table:exps_res}.\nextexperiment\label{table:exps_modellingissues})} & Observations          & \multicolumn{3}{l|}{Cache tag and set index (Def.~\ref{model:multi-way})}                                                                                                                                                                     \\ \cline{2-5} 
                     & Programs                                                                                & \multicolumn{3}{l|}{Random program generator}                                                                                                                                                                             \\ \cline{2-5} 
                     & \begin{tabular}[c]{@{}l@{}}Experiments\\ - Inconclusive\\ - Failure\end{tabular}        & \multicolumn{3}{l|}{\begin{tabular}[c]{@{}l@{}}22321\\ 0\\ 308\end{tabular}}                                                                                                                                              \\ \hline
\end{tabular}
 \caption{Invalidation of cache and faulty observational models.}
\label{table:exps_res}
\end{table}


\subsection{Partitioned cache observational model}
\label{sec:results:part_cache}
Next, we consider the partitioned cache observational model from Def.~\ref{model:partitioning}. 
That is, we partition 
the L1 cache of the Raspberry Pi 3 into two contiguous regions 
and assume that the attacker has only access to the second region.
Due to the prefetcher of Cortex-A53 we expect this model to be unsound
and indeed we could invalidate it.

To this end, we generated experiments for two variations of the model. Variation A splits the cache
at cache set 61, meaning that only cache sets 61--127 were considered accessible
to the attacker. Variation B splits the cache at cache set 64 (the midpoint),
such that cache sets 64--127 were considered visible.
The following program is one of the counterexamples for
variation A that have been discovered by \tool\ using the 
stride program generator of Section~\ref{sec:program-generation_strides}.

\begin{blstlisting}[hbox,enhanced,titlerule = 1pt,boxsep = 1pt,left = 10pt,right = 10pt,top=2pt,colframe = gray,center,parbox=false]{Program\hspace{4.1cm}Input 1\hspace{1.5cm}Input 2}
 ldr x2,  [x10, #0]    x10: 0x80100080   0x80100cc0
 ldr x20, [x10, #128]
 ldr x17, [x10, #256]
\end{blstlisting}

The counterexample exploits the fact that prefetching fills more lines than those
loaded by the program, provided the memory accesses happen in a certain stride
pattern. Thus, it essentially needs to have two properties: (i) two different
starting addresses for the stride, $a_1$ and $a_2$, with a cache set index that is
lower than 61 to avoid any observations in the model, and thus satisfying observational equivalence,
and (ii) one of $a_1$ and $a_2$ is close enough to the partition boundary.
In this case, automatic prefetching will continue to fill lines in subsequent sets,
effectively crossing the boundary into the attacker-visible region. 

In our experiments, we used a path guard to generate
only states that produce only memory accesses to the region of
the cache that is not visible by the attacker.
Additionally, we used term enumeration to force successive test cases to
start a stride on a different cache set and therefore cover the
different cache set indices. Without this guidance, the tool could
generate only experiments that affect the lower sets of the cache
and never explore scenarios that affect the sets with indices closer to the split boundary.

For variation B, we have not found such a
counterexample. The only difference is that the partition boundary is on line
64, which means that each partition fits exactly in a small page (4K). We
conjecture that the prefetcher does not perform line fills across small page
(4K) boundaries. This could be for performance reasons, as crossing
a page boundary can involve a costly page walk if the next page is not in the
TLB. If this is the case, it would seem that it is safe to use prefetching with
a partitioned cache, provided the partitions are page-aligned.
Table~\ref{table:exps_res}.\ref{table:exps_cachepart} summarizes our experiments for this model.


\subsection{Multi-way cache observational model}
In the remaining experiments, we consider the model of
Def.~\ref{model:multi-way} and we assume that
the attacker has access to the complete L1 cache.
Even if we expected this model to be sound, our experiments (Table~\ref{table:exps_res}.\ref{table:exps_cachetagandsetindex})
identified several counterexamples. We comment on two classes of
counterexamples below.


\smartparagraph{Previction}
Some counterexamples are due to an undocumented behavior that
we called ``previction'' because it causes a cache line
to be evicted before the corresponding cache set is 
full.
The following program compares \lstinline|x0| and \lstinline|x1| and executes a
  sequence of three loads. In case of equality, fourteen  
\lstinline|nop| are executed between the first two loads.

\begin{blstlisting}[hbox,enhanced,titlerule = 1pt,boxsep = 1pt,left = 10pt,right = 10pt,top=2pt,colframe = gray,center,parbox=false]{Program\hspace{3.9cm}Input 1\hspace{1.5cm}Input 2}
 cmp x0, x1            x0: 0x00000000   0x00000000
 b.eq #0x14            x1: 0x00000000   0x00000001
 ldr x9, [x2]          x2: 0x80100000   0x80100000
 ldr x9, [x3]          x3: 0x80110000   0x80110000
 ldr x9, [x4]          x4: 0x80120000   0x80120000
 b #0x48
 ldr x9, [x2]
 nop {14 times}
 ldr x9, [x3]
 ldr x9, [x4]
\end{blstlisting}

\lstinline|Input 1| and \lstinline|Input 2| are two states that exercise the two execution paths and 
have the same values for \lstinline|x2|, \lstinline|x3| and \lstinline|x4|, hence the
two states are observationally equivalent. Notice
that all memory loads access cache set 0.
Since the cache is 4-way associative and the cache is
initially empty, we expect no eviction to occur.

Executions starting in \lstinline|Input 2| behave as expected and 
terminate with the addresses of \lstinline|x2|, \lstinline|x3|, and \lstinline|x4| in
the final cache state.
However, the execution from \lstinline|Input 1| leads to a previction,
which causes the final cache state to only
contain the addresses of \lstinline|x3| and \lstinline|x4|. The address of
\lstinline|x2| has been evicted even if the cache set is not full.
Therefore the two states are distinguishable by the attacker.
Our hypothesis is that the processor detects a short sequence of loads to the same cache
set and anticipates more loads to the same cache set with no reuse of previously loaded
values. It evicts the valid cache line in order to make space for more
colliding lines.
We note that these cache entries are not dirty and thus eviction is most likely a
cheap operation. The execution of a \lstinline|nop| sequence probably
ensures that the first cache line fill is completed before the other addresses
are accessed.

\smartparagraph{Offset-dependent behaviors}
Our experiments identified further counterexamples that
invalidate the observational model. In particular, the following counterexample
also invalidates the observational model of
Def.~\ref{model:multi-way-pc}, where cache line offsets are not
observable.

\begin{blstlisting}[hbox,enhanced,titlerule = 1pt,boxsep = 1pt,left = 10pt,right = 10pt,top=2pt,colframe = gray,center,parbox=false]{Program\hspace{4.1cm}Input 1\hspace{1.5cm}Input 2}
 ldr x6, [x0]          x0 : 0x80108000   0x80108000 
 ldr x9, [x3, #4]      x3 : 0x800FFFFC   0x800FFFFC
 ldr x2, [x16]         x16: 0x80100020   0x80100000
 ldr x16,[x22]         x22: 0x8011FFF8   0x8011FFF8
 ldr x9, [x22,#8]  
\end{blstlisting}

This program consists of five consecutive load instructions. This program always produces
five observations consisting of the cache tag and set index of the five
addresses. \lstinline|Input 1| and \lstinline|Input 2| are observationally equivalent: they only
differ for \lstinline|x16|, which affects the address used for the third load,
but the addresses \lstinline|0x80100020| and \lstinline|0x80100000| have the same cache tag and
set index and only differ for the offset within the same cache line.
However, these experiments lead to two distinguishable
microarchitectural states. More specifically, execution from \lstinline|Input 1| results
in the filling of cache set 0, where
 the addresses of registers \lstinline|x0|, \lstinline|x3|,  \lstinline|x16| and
 \lstinline|x22 + 8| are present in the cache, while executions from \lstinline|Input 2|
 leads a cache state where the address of \lstinline|x0| is not in the
 cache and has been probably evicted.
This effect can  be the result of the interaction between cache
previction and cache bank collision~\cite{Osvik:2006:CAC:2117739.2117741,bernstein2005cache}, whose behavior
depends on the cache offset.
Notice that cache bank collision is undocumented for ARM Cortex-A53.
Tromer et al.~\cite{Tromer:2010:ECA:1713125.1713127} have shown that such offset-dependent behaviors can make insecure
side-channel countermeasures for AES that rely on making accesses to memory blocks
(rather than addresses) key-independent.

\subsection{Problems in model implementations}
Additionally to microarchitectural features that invalidate the formal
models, our experiments identified bugs of the 
implementation of the models: (1) the formalization of the ARMv8
instruction set used by the transpiler and
(2) the module that inserts BIR observation statements into
the transpiled binary to capture the observations that can be made
according to a given observational model.
Table~\ref{table:exps_res}.\ref{table:exps_modellingissues}
reports problems identified by the random program generator. Some of
these failing experiments result in distinguishable states while
others result in run-time exceptions. In fact, if the model
predicts wrong memory accesses for a program then our framework can generate test
inputs that cause accesses to unmapped memory regions.
The example program in Fig.~\ref{fig:proggen_example_randprog_cbnz}
exhibits both problems when executed with appropriate inputs.


\smartparagraph{Missing observations}
The second step of our framework translates binary programs to
BIR and  adds observations to reflect the observational
model under validation. In order to generate observations that
correspond to memory loads, we
syntactically analyze the right-hand side of BIR assignments. For
instance, for line $l2$ in Fig.~\ref{fig:bir_transpiled} we generate an observation
that depends on variable \lstinline|X1| because the expression of
assignment is \lstinline|LOAD(MEM, X1)|.
This approach is problematic when a memory load is immaterial
for the result of an instruction. For example, 
\lstinline|ldr xzr| and \lstinline|ldr wzr| instructions load from memory to
a register that is constantly zero.
The following
program loads from \lstinline|x30| into \lstinline|xzr|.

\begin{blstlisting}[hbox,enhanced,titlerule = 1pt,boxsep = 1pt,left = 10pt,right = 10pt,top=2pt,colframe = gray,center,parbox=false]{Program\hspace{3.3cm}Input 1\hspace{1.5cm}Input 2}
 ldr xzr, [x30]    x30: 0x80000040   0x800000038
\end{blstlisting}

The translation of this instruction is simply \lstinline|[JMP next_addr]|:
there is no assignment that loads from \lstinline|x30| because the register \lstinline|xzr|
remains zero.
Therefore, our model generates no
observations and any two input states are observationally
equivalent.
The ARM specification does not clarify that the microarchitecture can skip the
immaterial memory load. Our experiments show that this is not the
case and therefore our implementation of the model is not correct.
In fact, the program accesses cache set $\exidx{0x80000040} = 1$ for \lstinline|Input 1| and cache set
$\exidx{0x80000038} = 0$ for \lstinline|Input 2|, which results in distinguishable
states. Moreover, by not taking into account the memory access our
framework generates some tests that set \lstinline|x30| to unmapped
addresses and cause run-time exceptions.

\smartparagraph{Flaw in HOL4 ARMv8 ISA model}
Our tool has identified a bug of the  HOL4 ARMv8 ISA model. This
model has been used in several
projects~\cite{10.1145/3018610.3018621,DBLP:journals/jce/BaumannSD19}
as the basis for formal analysis and is used by our framework to
transform ARM programs to BIR programs. Despite
its wide adoption, we identified a problem in the semantics of instructions \textit{Compare and
  Branch on Zero} (CBZ) and \textit{Compare and Branch on
  Non-Zero} (CBNZ). These instructions implement a conditional jump based on the comparison
of the input register with zero. While CBZ jumps in case of
equality, CBNZ jumps in case of inequality. However, our tests identified that
CBNZ wrongly behaves as CBZ in the HOL4 model.


\section{Related work}
\label{sec:related}

\smartparagraph{Hardware Models}
Verification approaches that take into account the underlying hardware 
architecture have to rely on a formal model of that architecture. 
Commercial instruction set architectures (ISAs) are usually specified mostly 
in natural language, and their formalization is an active research direction. 
For example, Goel et al.~\cite{GoelHKG14} formalize the ISA of x86 in ACL2, Morrisett et 
al.~\cite{MorrisettTTTG12} model the x86 architecture in the Coq theorem 
prover, and Sarkar et al.~\cite{SarkarSNORBMA09} provide a formal semantics of 
the x86 multiprocessor ISA in HOL.
Moreover, domain-specific languages for ISAs have 
been developed, such as the L3 language~\cite{Fox12}, which has been used to 
model the ARMv7 architecture. 
As another example, Siewiorek et al.~\cite{Siewiorek82} proposed the 
\emph{Instruction-Set Processor} language for formalizing the semantics of the 
instructions of a processor. 

\smartparagraph{Processor Verification and Validation}
To gain confidence in the correctness of a processor model, it needs to be 
verified or validated against the actual hardware. This problem has received 
considerable attention lately. 
There are white-box approaches such as the formal verification 
that a processor model matches a hardware 
design~\cite{Fox03,BeyerJKLP06}. These approaches 
differ from ours in that they try to give a formal guarantee that a
processor model is a valid abstraction of the actual hardware, and to achieve 
that they require the hardware to be accessible as a white box.
More similar to ours are black-box approaches that validate an abstract 
model by randomly generated instructions or based on dynamic 
instrumentation~\cite{FoxM10,HouSTLH16}. 
Combinations of formal verification and testing approaches for 
hardware verification and validation have also been 
considered~\cite{bhadra2007survey}.

In contrast to our work, all of the approaches above are limited to 
functional correctness, and validation is limited to single-instruction test 
cases, which we show to be insufficient for information flow properties.
Going beyond these restrictions is the work of Campbell and 
Stark~\cite{CampbellS16}, who generate sequences of instructions as test 
cases, and go beyond 
functional correctness by including timing properties. Still, neither their 
models nor their approach is suitable to identify violations of information 
flow properties.

\smartparagraph{Validating Information Flow Properties}
To the best of our knowledge, we present the first automated approach to validate 
processor models with respect to information flow properties.
To this end, we build on the seminal works 
of McLean~\cite{McLean92} on non-interference, Roscoe~\cite{Roscoe95} on 
observational determinism, and Barthe et al~\cite{BartheDR11} on 
self-composition as a method for proving information flow properties.
Most closely related is the work by Balliu et al.~\cite{Balliu2014} on 
\textit{relational analysis} based on \textit{observational determinism}. 

These approaches are based on the different observational models that 
have been proposed in the literature.
For example, the program counter security model~\cite{Molnar2006} has been used when 
the execution time depends on the control flow of the victim.
Extensions of this model also make observable data
that can affect execution time of an instruction, or memory addresses
accessed by the program to model timing differences due to caching~\cite{AlmeidaBPV13}.

Many analysis tools use these observational models.
Ct-verif~\cite{Almeida2016} implements a sound 
information flow analysis by proving observational equivalence
constructing a product program.
CacheAudit~\cite{Koepf2015} quantifies information leakage 
by using abstract interpretation.

The risks of using unsound models for such analyses have been demonstrated by
the recent Spectre attack family~\cite{Kocher2018spectre}, which exploits
speculation to leak data through caches.
Several other architectural details require special caution when using
abstract models,
as some properties assumed by the models could be unmet. For instance,
cache clean operations do not always clean
residual state in implementations of replacement policies~\cite{Ge2016cflush}.
Furthermore, many processors do not provide sufficient means to close all leakage,
e.g., shared state cannot be cleaned properly on a context switch~\cite{Ge2016leaks}.
Finally, it has been shown that fixes relying on too specific assumptions can be circumvented by modifying the attack~\cite{SchaikGBR18}, and that attacks are possible even against formally verified software if the underlying processor model is unsound~\cite{GuancialeNBD16}. For these reasons, validation of formal models by directly measuring the hardware is of great importance.


\section{Concluding remarks}
\label{sec:conclusions}
We presented \tool{}, a framework for automatic validation of
observational models of side channels.
\tool{} uses a novel combination of symbolic execution, relational analysis, and
observational models to generate experiments.
We evaluated \tool{} on the ARM Cortex-A53 processor and we
invalidated all models of Sect.~\ref{sec:background:obsmodels},
i.e., those with observations that are cache-line-offset-independent.

Our results are summarized as follows:
(i) in case of cache partitioning,
the attacker can discover the victim's accesses to the other cache
partitions due to the automatic prefetcher;
(ii) the Cortex-A53 prefetcher seems to respect
4K page boundaries, like in some Intel processors;
(iii) a mechanism of Cortex-A53, which we called previction, can leak
the time between accesses to the same cache set;
(iv) the cache state is affected by the cache line offset of the
accesses, probably due to undocumented cache bank collisions like in some
AMD processors;
(v) the formal ARMv8 model had a flaw in the implementation of CBNZ;
(vi) our implementation of the observational model had a flaw in case
of loads into the constant zero register.
Moreover, since the microarchitectural features that lead to these findings 
are also available on other ARMv8 cores, including some that are affected by 
Spectre (e.g. Cortex A57), it is likely that 
similar behaviors can be observed on these cores, and that more powerful 
observational models, including those that take into account Spectre-like 
effects, may also be unsound.

These promising results show that \tool{} can
support the identification
of undocumented and security-relevant features of processors (like results (ii), (iii), and (iv))
and discover problems in the formal models (like
results (v) and (vi)).
In addition, users can drive test-case generation to conveniently explore
classes of programs that they suspect would lead to side-channel leakage (like
in result (i)). This process is enabled by path and term enumeration techniques
as well as custom program generators.
Moreover, \tool{} can aid vendors to validate
implementations with respect to
desired side-channel specifications.

Given the lack of vendor communication regarding security-relevant processor
features, validation of abstract side-channel models is of critical importance.
As a future direction of work, we are planning to extend
\tool{} for other architectures (e.g. ARM Cortex-M0 based
microcontrollers), 
noisy side channels (e.g. time and power consumption), and
other side channels (e.g. cache replacement state).
Moreover, we are investigating approaches to automatically repair an
unsound observational model starting from the counterexamples, e.g., by
adding state observations.
Finally, the theory in Sect.~\ref{sec:relation} can be used to
develop a certifying tool for verifying observational determinism.


\subsubsection*{\ackname}
We thank Matthias Stockmayer for his contributions to the symbolic execution
engine in this work. This work has been supported by the TrustFull project
financed by the Swedish Foundation for Strategic Research, the KTH CERCES Center
for Resilient Critical Infrastructures financed by the Swedish Civil
Contingencies Agency, as well as the German Federal Ministry of Education and
Research (BMBF) through funding for the CISPA-Stanford Center for Cybersecurity
(FKZ: 13N1S0762).

\bibliographystyle{splncs04}
\bibliography{main}
\appendix

\section{Program generation}
\label{app:proggen}
In this appendix, we present more details about the program generators used in \tool.

\subsection{Random generator}

Our first approach is based on \textit{randomly} generating test
programs. For this, we rely on the existing HOL4 model of ARMv8 ISA
and its instruction generator, which has been used to validate the functional
correctness of the model.
The generator splits the ARMv8 ISA into different classes, e.g.
\verb|Load-Store| and \verb|Conditional-Branch|. To generate an
instruction, it picks a class and randomly generates the binary encoding
of an instruction of the selected class. If the
generated binary is well-formed, it can be translated to the
corresponding ARM instruction mnemonic with randomly assigned source
and target registers. 

Since many instructions have no observable behavior, this approach may
generate programs which do not have any observation points and
therefore are not interesting for model validation (i.e., they could
not access memory).
In order to address this issue, we assign weights to instruction
classes and we modified the selection procedure to pick classes
accordingly. By increasing the weights of classes with observable
behaviors, we generate programs that are suitable to validate the
model. We also make sure that all jump instructions are directed
forward, thus naively guaranteeing termination of symbolic execution
of the programs, and that their target locations are within the
program address range,
thus preventing run-time exceptions.
Figure~\ref{fig:proggen_example_randprog_cbnz} shows one example of the programs
that this procedure generates.

\subsection{Monadic generators}

\tool\ supports \emph{monadic} program generators in the style of
QuickCheck.
In this context, a \emph{generator} is a first-class value that represents a
recipe to produce an assembly program based on a source of randomness.
%
This approach is based on a collection of simple generators
, which can be composed using a
monadic~\footnote{Monads are a standard pattern used to
  sequence effectful computations in functional programs. In our library, the
  side-effect hidden in the monad is the generation of random numbers.}
interface.
Type $\type{Gen}\,\type{A}$ represents generators of values of type $\type{A}$.
For example, $\term{arb\_addr} : \type{Gen}\,\type{int}$ generates a random
memory address.
Generators can be combined using sequencing and choice
operators, which allows a
generator to be chosen based on a desired distribution.
As a result, we obtain an embedded domain-specific language that we can use to
conveniently define generators of specific classes of programs that we are
interested in testing.
%



\smartparagraph{Load generator}
The simplest class of programs we support is programs that consist of a sequence
of arbitrary load instructions. Figure~\ref{fig:xldgen} shows the code for this
generator in our embedded language. We elide most of the definitions of the base
generators for brevity. The first line is an $\type{Operand}$ generator that
produces a load relative to a register, which is encoded as $\term{Ld}$ applied
to a pair of the form $(\term{off},\term{reg}) : \type{int} * \type{string}$,
where $\term{off}$ is the offset and $\term{reg}$ is the register name (elided).
Such pairs are made by the $\term{two}$ combinator, which takes two generators
$\term{g_1}$ and $\term{g_2}$ and produces a pair using $\term{g_1}$ and
$\term{g_2}$ to make the components.
These generators are: $\verb|elements [4,8,16]|$, which returns a random
element from the given list (in this example, either 4, 8 or 16); and
$\verb|arb_regname|$, which produces an arbitrary 64-bit register name. The
\verb|<$>| 
operator is used to add the constructor $\term{Ld}$ to the output of another
generator.
The following lines use similar primitives to make $\term{Load}$
instructions that load relative ($\verb|arb_load_rel|$) and immediate
($\verb|arb_imm|$) operands (respectively) into an arbitrary register
($\verb|arb_reg|$).
The $\verb|arb_load|$ generator chooses randomly between
$\verb|arb_load_rel|$ and $\verb|arb_load_abs|$ using the $\verb|oneof|$ choice
combinator.
Finally, the $\verb|arb_xld|$ generates a list of such loads using
the $\verb|arb_list_of|$ combinator, which makes lists with elements taken from
a given generator.
{
  \setlength{\belowcaptionskip}{-5pt}
\begin{figure}[t]
  \centering
  \begin{verbatim}
  val arb_ld = Ld <$> two (elements [4,8,16])
                          arb_regname;
  val arb_load_rel = Load <$> (two arb_reg arb_ld);
  val arb_load_abs = Load <$> (two arb_reg arb_imm);
  val arb_load = oneof [arb_load_rel, arb_load_abs];
  val arb_xld = arb_list_of arb_load;
\end{verbatim}\
  \vskip-10pt
  \caption{Load generator xld}
  \label{fig:xldgen}
\end{figure}
}

As a result, when we use $\verb|arb_xld|$ as a program generator we get programs
like the one in Figure~\ref{fig:proggen_examples} (first column).

\smartparagraph{Load strides}
\label{sec:program-generation_strides}
In order to experiment with cache prefetching, we have a specific generator that
makes programs with a memory access pattern that will trigger a prefetch. This
is achieved by making programs that always consist of a sequence of loads
relative to a fixed register $\term{src}$, of the form $\term{ldr}\quad
\term{r_i},\, \term{[src, offset_i]}$, where the $i$-th load offset is
$\term{offset_i}=(64*n*i)$ and $n \in [1..4]$ is the chosen stride step. The
target register $\term{r_i}$ is an arbitrary one for each load, but it is
selected such that it is different from the source register $\term{src}$.
Figure~\ref{fig:proggen_examples} (second column) shows an example of a program
generated in this way (with $n=1$).

\smartparagraph{Load sequences with branches}
\label{sec:program-generation_branches}
To explore potential counterexamples that may manifest the side channel in
the control flow of the program, we have a conditional branch generator that
produces structured if-statements.
The generator generates a comparison instruction (\texttt{cmp}) between two
arbitrary registers, and then a branch instruction (\texttt{b}) that jumps if
the compared registers are equal.
This generator is parametric, allowing the use of two arbitrary generators to
generate the code for each branch of the if-statement, say $b_1$ and $b_2$.
These blocks of code are then inserted after the comparison and branch, with a
suitable unconditional jump after $b_1$ to skip $b_2$. In pseudo-code, the
program would be of the form \verb|if reg1 = reg2 then b1 else b2|.

We use this conditional branch generator along with the straight-line load
generator to make programs that branch and perform different memory loads in
each branch. Figure~\ref{fig:proggen_examples} (third column) shows an example
of a program generated in this way.

\end{document}